\documentclass[10pt,a4paper]{article}
\usepackage{authblk}
\usepackage[utf8]{inputenc}
\usepackage{amsmath}
\usepackage{amsfonts}
\usepackage{amssymb}
\usepackage{graphicx}
\usepackage{dsfont}
\usepackage{float}
\usepackage{changepage}
\usepackage{cite}
\usepackage[left=2.5cm,right=2.5cm,top=2cm,bottom=2cm]{geometry}

\title{Deterministic Shaping of Quantum Light Statistics}

\author[1,2]{Garrett D. Compton}
\author[1]{Mark G. Kuzyk}

\affil[1]{Department of Physics and Astronomy, Webster Hall, Washington State University,
Pullman, WA 99164-2814}
\affil[2]{garrett.compton@wsu.edu}
\date{March 8, 2024}                     %% if you don't need date to appear
\begin{document}
\maketitle
\begin{abstract}
We propose a theoretical method for the deterministic shaping of quantum light via photon number state selective interactions. Nonclassical states of light are an essential resource for high precision optical techniques that rely on photon correlations and noise reshaping. Notable techniques include quantum enhanced interferometry, ghost imaging, and generating fault tolerant codes for continuous variable optical quantum computing. We show that a class of nonlinear-optical resonators can transform many-photon wavefunctions to produce structured states of light with nonclassical noise statistics. The devices, based on parametric down conversion, utilize the Kerr effect to tune photon number dependent frequency matching, inducing photon number selective interactions. With a high amplitude coherent pump, the number selective interaction shapes the noise of a two-mode squeezed cavity state with minimal dephasing, illustrated with simulations. We specify the requisite material properties to build the device and highlight the remaining material degrees of freedom which offer flexible material design. 
\end{abstract}

\section{Introduction}

Systems that utilize quantum information resources generally take advantage of coherent macroscopic superposition and/or entanglement. The increased information density and sensitivity to measurements of such quantum states, compared to their classical counter parts, is central to their application in modern quantum information technologies like quantum key distribution\cite{sango10.01, brask10.01}, quantum computing \cite{sakag23.01, gotte01.01, lidar98.01}, and metrology at the fundamental noise limit\cite{polin20.01, oszma16.01, unter18.01}. Efforts toward engineering the spatio-temporal profile of quantum states of light have already found great success\cite{vasi10.01, morin19.01, huang23.01}. This article focuses instead on shaping the noise statistics of quantum states of light for such applications. Both macroscopic superposition and entanglement are generically found in states supported by nonlinear systems \cite{yurke86.01, koles23.01}. However, optical nonlinearities are typically weak for closed systems. Much attention has therefore been paid to developing methods for generating quantum states of light which utilize wavefunction collapse as a source of nonlinear evolution, where a particular measurement outcome heralds the desired quantum state \cite{asava17.01, takase21.01, chen23.01, fukui22.01, hayun21.01}. Though hugely successful, heralded state preparation is inherently non-deterministic, making sequential preparations low probability events.

The approach presented here overcomes low heralding success rates using the paradigm of deterministic generation of quantum states. Examples in the literature can be found where the probabilistic nature of measurement is circumvented\cite{beige00.01, gonz15.01, verst09.01, vertc23.01, camac22.01, ribei21.01, hu19.01, imamo97.01, gonz15.01, porra08.01, perar20.01, verst09.01, law96.01, krast15.01, heere15.01}.

One example utilizes the quantum Zeno effect \cite{beige00.01, gonz15.01, verst09.01}. By strongly coupling a specially engineered system to an environment, the state of the system is continuously projected to a subspace of states that are effectively uncoupled from the environment, called a decoherence free subspace (DFS). External driving fields may then evolve states deterministically and coherently within the DFS.

With the advent of epsilon-near-zero materials \cite{vertc23.01} and polariton enhanced scattering, commonly facilitated by electronically induced transparency, \cite{camac22.01, ribei21.01, hu19.01, imamo97.01}, large effective nonlinearities become increasingly realistic as well. When realized in a high-Q optical resonator, strong optical nonlinearities with relatively long evolution time and low loss facilitate deterministic quantum optical state engineering. Trapped atom and other quantum emitter systems show great promise for this application  {\cite{gonz15.01, porra08.01, perar20.01, verst09.01, law96.01, krast15.01, heere15.01}}.

Such systems accomplish universal control of an oscillator state by relying on an array of emitters or by coupling to a single two level emitter. With an array of quantum emitters, the quantum information content of a state of interest is first written onto the array, typically in a decoherence free subspace, then mapped to super-radiant light-matter states which couple to an optical waveguide \cite{gonz15.01, porra08.01, perar20.01, verst09.01}. Alternatively, arbitrary state transformations may be realized in a cavity with a single quantum emitter by repeatedly addressing the emitter with a control beam to introduce a Berry phase to cavity field Fock states and applying coherent displacement operations to the cavity field in sequences prescribed within {\cite{law96.01, krast15.01, heere15.01}}. Notably, Krastanov et al. {\cite{krast15.01, heere15.01}} utilize a dispersive coupling between the emitter and a microwave cavity field to realize a selective number-dependent arbitrary phase (SNAP) gate which selectively rotates the phase of each Fock state of the cavity field.

This work introduces an alternative technique for precision shaping of quantum noise statistics without the need for precise emitter engineering and addressing. We consider a material with first, second, and third order electric susceptibilities at optical frequencies \footnote{It is straightforward to generalize our approach for nonperturbative constitutive relations.} and show that when second and third order nonlinearities are comparable in strength, the optical Kerr effect alters the frequency matching condition for optical parametric oscillation such that the interaction between field states is strongly photon number dependent. Our analysis separates the phase rotating interactions from the mode mixing interactions, providing us a deeper intuition for the dynamics we observe. Specifically, we first solve the combined first and third order system exactly and study the second order mode mixing in the interaction picture.  The general dynamical characteristics are illustrated and we utilize the effect to demonstrate a procedure for deterministic generation of approximate Fock states, among other interesting states. General material constraints are highlighted and developed into figures of merit to guide experimental realizations of this method.
%We illustrate a procedure for generating a mixed amplitude Bell state of Schr\"odinger cat states

\section{Background}
We begin with a brief discussion of electromagnetic field quantization. Dirac recognized that Hamiltonian mechanics has simple rules for canonical quantization, that Lagrangian mechanics is well suited for relativistic gauge invariant classical field theories, and that they both poorly handle what the other does well. So, to formulate relativistic quantum theory, he generalized Hamiltonian mechanics to handle systems with constraints\cite{dirac50.01}. His mechanics encodes symmetries that are simple to write into a Lagrangian, into constraints on a Hamiltonian, recovering Hamilton's equations of motion with a generalized Poisson bracket called the Dirac bracket. Canonical quantization of Dirac's bracket reveals, in part, Dirac's relativistic theory of quantum mechanics. Born and Infeld adopted Dirac's procedure  to show that Maxwell's equations in the Coulomb gauge assume a Hamiltonian formulation \cite{born34.02,born34.01} where the Hamiltonian is the classical field energy

\begin{equation}
H = \int_V dv\; \int_{-\infty}^t dt\; \frac{E\cdot \partial_{t'} D }{2} + \frac{H\cdot \partial_{t'} B}{2}  \label{energy}
\end{equation}

 and the fields obey a version of the canonical Dirac bracket relations,

\begin{equation}
\begin{split}
&\{B^j(r), B^k(r')\} = \{D^j(r), D^k(r')\} = 0 \\
&\{D^j(r),B^k(r')\} = \epsilon^{jlk}\frac{\partial}{\partial x^l}\delta(r-r') \equiv \delta^{tr}(r-r') \;,
\end{split}
\end{equation}

with the second equation defining the transverse delta function $\delta^{tr}(r-r')$. \footnote{The transverse delta function captures the constraints on the field enforced by the Coulomb gauge. In particular, note that in a medium without free charges or free currents, Maxwell's equations ensure $D$ and $B$ are always transverse to the direction of radiation propagation, unlike $E$ and $H$. Furthermore, since both $D$ and $B$ are divergenceless, so must be their commutator. The transverse delta function is consistent with these two properties. It has no divergence and acts as an ordinary delta function on transverse fields.}According to Dirac's canonical quantization, the field amplitudes are promoted to operators, and the Dirac brackets become commutators

\begin{equation}
\{D^j(r),B^k(r')\} = \delta^{tr}(r-r')  \rightarrow  [D^j(r),B^k(r')] = i\hbar \delta^{tr}(r-r') \;.
\end{equation}

Quesada et al. \cite{quesa17.01} later showed that, to correctly recover Maxwell's equations in nonlinear dielectric media from the Born-Infeld Dirac brackets, it is imperative to consider the displacement field and magnetic flux density to be the fundamental dynamical variables. Here 'fundamental' indicates that $D$ and $B$ are Hermitian linear combinations of Bose operators, i.e. they are Boson fields, whereas the electric and magnetic fields $E$ and $H$ are generally higher polynomials of Bose operators\footnote{For a rigorous discussion see \cite{quesa17.01}. Quesada provides the following intuitive argument for why neither $E$ nor $H$ can be fundamental: Suppose the Hamiltonian is order $N+1$ in Bose operators. If $E$ is fundamental, it is of order 1. If $B$ is order 1, then $\partial_t B \sim [B,H] \sim \nabla \times E$ is order N. Thus $E$ is order N, a contradiction. A similar argument holds for $H$. No contradiction occurs with $D$ and $B$ as fundamental. }. For modern derivations and perspectives of quantum field theory in linear dielectric media, see \cite{bhat06.01, rayme20.01, hille84.01, drumm14.01}.

For a finite linear cavity, the eigenstates of the system are enforced by the dielectric structure and boundary conditions of the cavity. Let us restrict ourselves to one effective spatial dimension, scalar susceptibilities, and a single polarization of the field modes without loss of generality\footnote{Our analysis depends only on the algebraic structure of the Hilbert space and the dynamics generated by the Hamiltonian, so is easily generalized to richer mode structures.}. This amounts to a single parameter family of eigenstates. Consider a uniform cavity of length $L$ with periodic boundary conditions, as is the case for a microring resonator  \cite{quesa22.01}. The allowed wavenumbers, $k$, within the cavity are integer multiples of $\frac{2\pi}{L}$. The fields $D$ and $B$ are physical, therefore real valued. Each may then be expanded with respect to the basis of eigenmodes in the following form,

\begin{equation}
\begin{split}
&D(z) = \sum_k D_k(z)  = \frac{1}{\sqrt{AL}}\sum_k \alpha_k e^{ikz} + \alpha_k^* e^{-ikz}\\
&B(z) = \sum_k B_k(z)  =  \frac{1}{\sqrt{AL}}\sum_k \beta_k e^{ikz} + \beta_k^* e^{-ikz}\;,
\end{split}
\end{equation}

where $A$ represents the transverse cross sectional area of the modes in the cavity\footnote{If the refractive index is not isotropic, $A$ defines an effective cross sectional area, discussed in \cite{quesa22.01}.}.

The field amplitudes of a single photon, $\zeta_k$ -- defined later, are determined such that the energy of a photon with well defined momentum $\hbar k$ agrees with the Einstein energy of a photon, $\hbar \omega(k)$. Suppose a dispersive permitivity such that the Fourier components of the linear electric field and displacement field are related by $\tilde E(k) = \frac{1}{\epsilon(k)} \tilde D(k)$. If the medium is dispersive, each frequency mode solves a different eigenvalue equation, so normalization requires special care  \cite{rayme20.01}. We rescale the amplitude by the single photon amplitude, $\alpha_k \rightarrow \zeta_k a_k$, and promote the amplitude variable to a Bose operator obeying the canonical commutation relations

\begin{equation}
[a_{k}, a_{k'}] = 0 , \quad [a_{k}, a_{k'}^\dagger] = \delta_{kk'}\;.
\end{equation}

The displacement field operator becomes

\begin{equation}
D(z) = \sum_k i \sqrt{\frac{\hbar \omega(k) v_{g,k}n_k^3}{2ALc}}\Big(a_ke^{ikz} - a^\dagger e^{-ikr}\Big) \equiv \sum_k i \zeta_k\Big(a_ke^{ikz} - a^\dagger e^{-ikr}\Big)\;, \label{zeta}
\end{equation}

where we have adopted a common quantum optics phase convention of initial phase $\pi/2$, $ v_{g,k}$ is the group velocity, and $n_k$ the refractive index of the mode with wave-vector $k$. The magnetic flux field operator obtains a similar form\cite{rayme20.01}.

Having properly dealt with linear dispersion, the Hamiltonian for linear quantum optics takes the form of an infinite set of harmonic oscillators

\begin{equation}\label{H1}
 H^{(1)} = \sum_k \hbar \omega(k) a_k^\dagger a_k\;.
\end{equation}

\subsection{Nonlinear Quantum Optics}
We introduce nonlinear optics assuming that the dispersion of nonlinear susceptibilities is negligible for frequency bands we are interested in.
The electric field is related to the displacement field and polarization field, $P$, via \cite{jacks03.01}

\begin{equation}
E = D - 4\pi P = \frac{1}{\epsilon} D -4\pi P^{nl}\;.
\end{equation}

Recall from Equation \eqref{energy} the electric field contribution to the field energy, $\frac{E\cdot D}{2}$. The energy term $\frac{1}{\epsilon} D\cdot D$ is handled in the linear part of the Hamiltonian upon quantization. The remaining interaction $-P^{nl}\cdot D$ relays the effect of the nonlinear polarization $P^{nl}$.

We assume that the polarization at a point in space depends on the displacement field at that point, i.e. the response function is local.  Expanding the nonlinear polarization in a power series of the displacement field amplitudes up to third order, the part of the Hamiltonian associated with classical nonlinear dynamics is

\begin{equation}
\begin{split}
    H^{(nl)} &= \frac{-P^{nl}\cdot D}{2}  = \\&  -2\pi
    \sum_{k_0,k_1,k_2}\int dr^3\;\Gamma^{(2)}(\vec{r};\omega(k_0), \omega(k_1),\omega(k_2)))D_{k_0}(\vec{r},t)D_{k_1}(\vec{r},t)D_{k_2}(\vec{r},t) \\& -2\pi
   \sum_{k_0,k_1,k_2,k_3}\int dr^3\;\Gamma^{(3)}(\vec{r};\omega(k_0), \omega(k_1),\omega(k_2),\omega(k_3)))
   \\&
   \times D_{k_0}(\vec{r},t)D_{k_1}(\vec{r},t)D_{k_2}(\vec{r},t)D_{k_3}(\vec{r},t) \equiv H^{(2)} + H^{(3)} ,
\end{split}\label{Hnl}
\end{equation}

where $\Gamma^{(2)}$ is the second order displacement field susceptibility that is responsible for optical parametric oscillation (OPO) -- often referred to as parametric down conversion. $\Gamma^{(3)}$ is the third order displacement field susceptibility responsible for four effects: the two optical Kerr effects called self phase modulation (SPM) and cross phase modulation (XPM), a mode mixing process called four wave mixing (FWM), and a static field induced OPO. Both are defined by the power series expansion of the polarization field in terms of displacement field \cite{quesa22.01}.

The ordinary nonlinear susceptibilities are defined analogously for the electric field \cite{boyd20.01}, namely

\begin{equation}\label{PExpandDandE}
 P = \Gamma^{(1)}D + \Gamma^{(2)}D^{\otimes 2} + \Gamma^{(3)}D^{\otimes 3} + ... = \chi^{(1)}E + \chi^{(2)}E^{\otimes 2} + \chi^{(3)}E^{\otimes 3} + ...\;
\end{equation}

Substituting the relation $D = E+4\pi P(E)$ into Equation \eqref{PExpandDandE} and collecting terms with common powers of electric field, ignoring tensor structure and dispersion, yields the relations

\begin{equation}\
\begin{split}
\Gamma^{(1)} = \frac{\chi^{(1)}}{\epsilon},\quad
\Gamma^{(2)} = \frac{\chi^{(2)}}{\epsilon^3},\quad
 \Gamma^{(3)} = \frac{\chi^{(3)}}{\epsilon^4} - 8\pi \frac{(\chi^{(2)})^2}{\epsilon^5}\;.
\end{split}\label{gamma2chi}
\end{equation}

\section{Approach}\label{approach}
We aim to show how strong Kerr nonlinearities affect OPO. The essential result is that photon number dependent frequency shifts from the Kerr effect generate a photon number dependent detuning in OPO. The number dependent detuning is exploited for coherent deterministic number selective shaping of noise statistics in select modes. To make this effect clear, we constrain our view to the subspace of three optical modes engaging in OPO, called signal, idler, and sum\footnote{In general, the process we discuss can occur over the spectrum of cavity modes. This will be addressed in future papers.}. We first solve the system exactly under the action of the sum of Hamiltonian terms that preserve photon number in each mode. Evolution under this Hamiltonian gives nontrivial structure to the time dependent displacement field operator; it is scaled by a diagonal unitary operator with a frequency that is a function of the photon number operators. OPO is analyzed in the interaction picture, exhibiting photon number selective behavior. The nature of this behavior is discussed, and explored numerically in Section \ref{sims}.

\subsection{Eigenstates of Kerr Hamiltonian}
The Hamiltonian $H^{(3)}$ from Equation \eqref{Hnl} is responsible for two distinct photon number density preserving processes, SPM and XPM, and a mode mixing process, FWM. FWM is a rich effect, leading to optical frequency combs and soliton propagation in micro-resonators \cite{pasqu18.01}. However, our aim is to study an entirely different phenomenon in the same environment, so we do not wish to handle these complications presently. We will suppose the dispersion relation of our system is specified such that no FWM is phase matched for the active modes.

Unlike FWM, SPM and XPM are always phase matched. These effects, together with the linear Hamiltonian, dominate the phase dynamics of the system. The corresponding Hamiltonian is diagonal in the Fock basis and, equivalently, conserves photon number for each mode. Our present objective is to solve for the time evolution of the field operator under photon number density preserving Hamiltonian $H_0 \equiv H^{(1)} + H^{(3)}$ such that $H = H_0 + H^{(2)}$, where $H^{(1)}$, $H^{(2)}$, $H^{(3)}$ are defined in equations \eqref{H1} and \eqref{Hnl}. We later show how this time evolution affects $H^{(2)}$, producing a photon number selective interaction under the right conditions. We will consider the dynamics of three modes interacting in the cavity; a sum, signal, and idler mode with frequencies $\omega_\Sigma, \omega_s, \omega_i$ respectively.

We assume permutation invariance of all susceptibilities\footnote{Permutation invariance typically does not hold near resonance, where we expect our system to reside given the constraints derived later. However, accounting for variability in $\Gamma^{(3)}$ simply amounts to substituting the average $\bar \Gamma^{(3)}$ in place of $\Gamma^{(3)}$ and does not change the results.} and abbreviate the frequency dependence of Kerr susceptibilities $\Gamma^{(3)}(-\omega_1, \omega_1,-\omega_2,\omega_2)) = \Gamma^{(3)}_{1,2} = \Gamma^{(3)}_{2,1}$. Furthermore, we uphold the rotating wave approximation\footnote{The rotating wave approximation is synonymous with the secular approximation.} and assume spatially constant third order susceptibility. After ordering all permutations of the creation and annihilation operators to be simple powers of the mode number operators $N_j = a_j^\dagger a_j$, the Hamiltonian $H_0$ pertaining to the sum, signal, and idler modes is

\begin{equation}\label{H0}
\begin{split}
H_0 &= \Big(\hbar \omega_\Sigma -12\pi L \zeta_{\Sigma}^2(2\zeta_{\Sigma}^2\Gamma^{(3)}_{\Sigma, \Sigma} + \zeta_{s}^2\Gamma^{(3)}_{\Sigma,s} + \zeta_{i}^2\Gamma^{(3)}_{\Sigma, i})\Big)N_\Sigma\\&
+\Big(\hbar \omega_s -12\pi L \zeta_{s}^2(2\zeta_{s}^2\Gamma^{(3)}_{s, s} + \zeta_{i}^2\Gamma^{(3)}_{s,i} + \zeta_{\Sigma}^2\Gamma^{(3)}_{s,\Sigma})\Big)N_s\\&
+\Big(\hbar \omega_i -12\pi L \zeta_{i}^2(2\zeta_{i}^2\Gamma^{(3)}_{i,i} + \zeta_{\Sigma}^2\Gamma^{(3)}_{i,\Sigma} + \zeta_{s}^2\Gamma^{(3)}_{i,s})\Big)N_i\\&
-24\pi L\Big(\zeta_\Sigma^4\Gamma^{(3)}_{\Sigma, \Sigma}N_\Sigma^2 + \zeta_s^4\Gamma^{(3)}_{s,s}N_s^2 + \zeta_i^4\Gamma^{(3)}_{i,i}N_i^2 \\&
\quad \quad \quad \quad + \zeta_\Sigma^2\zeta_s^2\Gamma^{(3)}_{\Sigma, s}N_\Sigma N_s + \zeta_\Sigma^2\zeta_i^2\Gamma^{(3)}_{\Sigma, i}N_\Sigma N_i + \zeta_i^2\zeta_s^2\Gamma^{(3)}_{i, s}N_iN_s\Big)
\end{split}
\end{equation}

Heisenberg's equation of motion, $\dot a = \frac{1}{i\hbar}[a,H_0]$, tells us the time evolution for the field annihilation operators $a_\Sigma, a_s, a_i$.

To condense our notation, let us define the Kerr frequencies

\begin{equation}
g_{ij} \equiv \frac{24\pi AL}{\hbar} \zeta_i^2\zeta_j^2\Gamma^{(3)}_{ij}\label{Kerrfreq}
\end{equation}

where $g_{jj}$ are the frequencies of self phase modulation and $g_{i,j\neq i}$ are the frequencies of cross phase modulation. These frequencies are proportional to the classical Kerr frequency shifts $\delta\omega$ in accordance with the relation $g\langle N\rangle  = \delta\omega I$ where $I$ is the classical intensity and $\langle N\rangle $ is the mean photon number. The time evolution of each field annihilation operator under $H_0$ is

\begin{equation}
\begin{split}
a_\Sigma(t) &= \exp\Big\{-i\Big(\omega_\Sigma - (g_{\Sigma\Sigma} + \frac{g_{\Sigma s}}{2} + \frac{g_{\Sigma i}}{2}) - g_{\Sigma\Sigma}(2N_\Sigma + 1) + g_{\Sigma s}N_s + g_{\Sigma i}N_i)\Big)t\Big\}a_\Sigma(0)\\
a_s(t) &= \exp\Big\{-i\Big(\omega_s - (g_{ss} + \frac{g_{s\Sigma}}{2} + \frac{g_{si}}{2}) - g_{ss}(2N_s + 1) + g_{s\Sigma}N_\Sigma + g_{s i}N_i)\Big)t\Big\}a_s(0)\\
a_i(t) &= \exp\Big\{-i\Big(\omega_i - (g_{ii} + \frac{g_{i\Sigma}}{2} + \frac{g_{is}}{2}) - g_{ii}(2N_i + 1) + g_{i\Sigma}N_\Sigma + g_{is}N_s)\Big)t\Big\}a_i(0)\;.
\end{split}\label{a(t)}
\end{equation}
Note that the exponentials above do not commute with the $t=0$ field annihilation operators. To further simplify our notation we call $a(0) = a$, and use $\xi$ to represent the exponential operators on the right hand side of Equation \eqref{a(t)} such that
\begin{equation}
 a_\Sigma(t) = \xi_\Sigma a_\Sigma, \quad a_s(t) = \xi_s a_s, \quad a_i(t) = \xi_i a_i\label{timedep}
\end{equation}
Substituting these operators into Equation \eqref{zeta} yields the time dependent displacement field operator for the nonlinear field described by $H_0$,
\begin{equation}
D_k(z,t) = i\zeta_k\Big(\xi_k(t)a_ke^{ikz} - e^{-ikz}a_k^\dagger \xi_k^*(t)\Big)\;.\label{D(t)}
\end{equation}
The time dependent displacement field operator is the essential result of this subsection and the starting point for our analysis in the following subsection {\ref{NSOPO}}. Let us take a brief detour to understand how the $\xi$ operators affect optical  pumping and derive our first constraint.
\subsubsection{Pump photon blockade}
The $\xi$ operators neatly quantify the photon blockade effect, reviewed in \cite{imamo97.01}, which affects optical pumping of the resonator. To see this, consider a cavity pumped by a near monochromatic beam that is near resonance to a cavity mode with wavenumber $k_j$. As the photon number in the cavity increases from $N_j$ to $N_j + 1$, the frequency shifts by $g_{jj}(2N_j + 1)$. If the frequency of the next excitation shifts out of the pump line width, the cavity will no longer be excited, i.e. additional photons are blockaded from entering by the photons already in the cavity. Consequently, obtaining a high cavity intensity requires a pump beam with a wide linewidth in the presence of very strong Kerr nonlinearities. However, a wide pump line width yields unwanted excitations, so it is best for the Kerr nonlinearities involving the pump frequency to be small.

If small Kerr frequencies at the pump frequency are challenging to achieve, we consider the following constraint: Self phase modulation by a positive susceptibility red shifts the frequency by $-g_{jj}N_j^2$. If $\Gamma^{(3)}_{jj}>0$, the largest photon number achievable in the cavity pumped by a beam with a spectral edge at some fraction $f$ of $\omega_j$ is
\begin{equation}
 N_{MAX} = \sqrt{\frac{(1-f)\omega_j}{g_{jj}}}\label{Nmax}
\end{equation}

When considering optical parametric oscillation in the next sections, we drive the sum frequency cavity mode with an external pump. It is most important that the external pump spectrum  does not overlap with the signal and idler modes to avoid unwanted excitation. Therefore it is necessary that $f>1/2$ and preferable that $f>3/4$. We discuss later in Section \ref{Purity}, and simply note now, that a large coherent state for the sum mode minimizes dephasing and enhances OPO efficiency. Because unwanted excitations and high pump intensity are coincident, it is best for the Kerr nonlinearities involving the pump frequency to be negative, if not small.

\subsection{Number Selective Optical Parametric Oscillation}\label{NSOPO}
The previous section solved the Kerr Hamiltonian exactly and introduced the operators $\xi$ to simplify our analysis of optical parametric oscillation. With an exact time evolution for the Kerr photons in hand, we construct the Hamiltonian for number selective optical parametric oscillation (NSOPO).  In the interaction picture, the OPO Hamiltonian evolves via the photon number preserving propagator, $U_0 = \exp\{-iH_0t/\hbar\}$. From Equation \eqref{Hnl} we obtain the time dependent OPO Hamiltonian in the interaction picture,

\begin{equation}
 H_I = U_0^\dagger H^{(2)}U_0 = -2\pi\sum_{k_0,k_1,k_2}\int dr^3\;\Gamma^{(2)}(\vec{r};\omega(k_0), \omega(k_1),\omega(k_2)))D_{k_0}(\vec{r},t)D_{k_1}(\vec{r},t)D_{k_2}(\vec{r},t)\;.\label{HIfull}
\end{equation}

We assume the susceptibility $\Gamma^{(2)}$ is spatially local, spatially constant, and frequency permutation invariant\footnote{As is the case for $\Gamma^{(3)}$, permutation invariance is not valid for the strong nonlinearities required of NSOPO.  However, accounting for variability simply amounts to substituting the average $\bar \Gamma^{(2)}$ in place of $\Gamma^{(2)}$ and does not change the result.}. The time dependent displacement field operators, Equation \eqref{D(t)}, are substituted into Equation \eqref{HIfull}. Terms of the form $a^\dagger_\Sigma a^\dagger_i a^\dagger_s$ and $a_\Sigma a_sa_i$ are neglected by assuming the rotating wave approximation for the three modes, but the time dependence from the Kerr induced frequency mismatch is preserved. All possible permutations of the three relevant mode contributions to the field operators are reordered to a common tractable form, accounting for the noncommutativity of $a$ and $\xi$. The simplified interaction Hamiltonian is thus

\begin{equation}
 H_I = -i12\pi AL\zeta_\Sigma\zeta_s\zeta_i \left(\Gamma^{(2)}_{\Sigma, s, i}a^\dagger_\Sigma a_sa_i\xi_\Sigma^*\xi_s\xi_ie^{-i(2g_{ss} + 2g_{ii} + g_{si} - g_{\Sigma s} - g_{\Sigma i})t} - h.c. \right)
\end{equation}

where h.c. denotes hermitian conjugate.

Suppose the sum frequency mode is a high amplitude coherent state $|\alpha_\Sigma\rangle$ such that $\bar N_{\Sigma} = |\alpha_\Sigma|^2 \gg \bar N_s, \bar N_i$. Then we can reasonably replace the operator $a_\Sigma$ with the c-number $\alpha_\Sigma$ without worrying much about dephasing effects. For example, if shaping of signal and idler modes is done in a range of $m$ photons, the purity of the manipulated quantum state is on the order of $e^{-|\frac{m}{2\alpha_\Sigma}|^2}$\footnote{See
Section \ref{Purity}}.

To simplify notation we define the NSOPO complex squeezing parameter $\Xi$ and OPO resonance detuning frequency $\Delta(N_\Sigma, N_s, N_i)$ such that

\begin{equation}
\begin{split}
&\Xi e^{i\Delta t} \equiv \Big(\frac{12\pi AL}{\hbar}\zeta_\Sigma\zeta_s\zeta_i\Gamma^{(2)}_{\Sigma, s, i}\alpha_\Sigma^*\Big)\Big(\xi_\Sigma^*\xi_s\xi_ie^{-i(2g_{ss} + 2g_{ii} + g_{si} - g_{\Sigma s} - g_{\Sigma i})t}\Big)
\end{split}
\end{equation}

where

\begin{equation}
\Xi \equiv \frac{12\pi AL}{\hbar}\zeta_\Sigma\zeta_s\zeta_i\Gamma^{(2)}_{\Sigma, s, i}\alpha_\Sigma^*\;,\label{Xi}
\end{equation}

\begin{equation}
\begin{split}
\Delta \equiv \delta+ G_\Sigma N_\Sigma + G_S  (N_s + 1) + G_i (N_i + 1)\;,
\end{split}
\end{equation}

and

\begin{equation}
\begin{split}
 &\delta \equiv \omega_\Sigma - \omega_s - \omega_i + 2(g_{ss} + g_{ii} - g_{\Sigma \Sigma})\\
 &G_\Sigma \equiv g_{\Sigma s} + g_{\Sigma i} - 2g_{\Sigma \Sigma}\\
 &G_s \equiv  2g_{ss} + g_{is} - g_{\Sigma s}\\
 &G_i \equiv 2g_{ii} + g_{si} - g_{\Sigma i}
\end{split}\label{KerrDetuneings}
\end{equation}

are the set $\{G\}$ of detuning weights for each photon number, as well as a static detuning $\delta$.

The interaction Hamiltonian is now expressed compactly as

\begin{equation}
H_I = -i\hbar\Big(a_sa_i\Xi e^{i\Delta t} - e^{-i\Delta t}\Xi^*a_i^\dagger a_s^\dagger\Big)
\end{equation}

Without Kerr nonlinearities, non-degenerate parametric down conversion generates 2-mode squeezing between the signal and idler modes \cite{alam18.01}. With Kerr nonlinearities, the squeezing parameter hosts a detuning oscillation with photon number dependent frequency, making two mode squeezing amplitude dependent. Let us explore precisely how this feature affects the interaction.

The OPO detuning $\Delta(N_\Sigma, N_s, N_i)$ governs how number states experience 2-mode squeezing. A simple, and very useful case to study is when a large-$\alpha_\Sigma$ coherent state $|\alpha_\Sigma\rangle$ is injected into the nonlinear cavity in the presence of a 2-mode squeezed state of signal and idler frequencies. Experimentally this would also require a method for coherently injecting a 2-mode squeezed state into the nonlinear cavity initially in the vacuum state. Acknowledging the technological challenge\footnote{This may be realized, for example, with an acousto-optic modulator. }, we will assume it is possible and investigate its consequences\footnote{This example also captures the behavior of NSOPO via degenerate parametric down conversion and second harmonic generation. In these cases, the fundamental mode and second harmonic are entirely uncorrelated and coupled into the cavity without a need to preserve correlations. We choose to study NSOPO on a two mode squeezed state rather than a single mode state to highlight how noise shaping is identically present in the entangled signal and idler modes.}.

Assume an initial 2-mode squeezed state -- a state which is a superposition of all equal populations of signal and idler photons. Evident from the interaction operator $a_\Sigma^\dagger a_s a_i$, the difference in signal and idler photon numbers is a conserved quantity, so the state remains a superposition of equal photon numbers for all times much less than the cavity decay time. When $G_\Sigma = 0$, the OPO detuning $\Delta$ is independent of the sum frequency photon number and NSOPO is coherently driven by the sum frequency beam. Each OPO interaction event will decrement or increment the signal and idler photon numbers by one, transporting probability amplitude along the space of states. Adjacent states see a difference in detuning of

\begin{equation}
\gamma \equiv G_s + G_i = 2(g_{ss} + g_{ii} + g_{is} - g_{\Sigma\Sigma})\label{gamma}
\end{equation}

Suppose there is a photon number, $N_s = N_i = n'$ such that the detuning is minimal. We call $n'$ the shape center. Then the OPO Hamiltonian in Dirac notation takes the form

\begin{equation}
H_I = -i\hbar \sum_{m= 0}^\infty (m+1)\Big(\Xi e^{i(\gamma (n'-m) + \Delta(n'))  t} |m\rangle \langle m+1| - \Xi^* e^{-i(\gamma (n'-m) + \Delta(n')) t} |m+1\rangle \langle m|\Big).\label{HI}
\end{equation}

This Hamiltonian generates particularly interesting evolution when $\Xi$ and $\gamma$ are of similar magnitude, however challenging to describe analytically. We explore this behavior in the following section.

\section{Results and Discussions}
In the previous section we developed a theoretical method for directly targeting photon number states to coherently shape the quantum statistics of an optical field in a cavity. In the following we numerically simulate the model and illustrate how the parameters can be controlled to generate quantum states with useful structure. The constraints established here and in Section \ref{approach} are elaborated on and brought together to determine what regions of parameter space are viable for experimental realization.

\subsection{Simulations}\label{sims}
\begin{figure}[H]
\begin{adjustwidth}{-1cm}{-1cm}
\centering
\includegraphics[width=17.5cm]{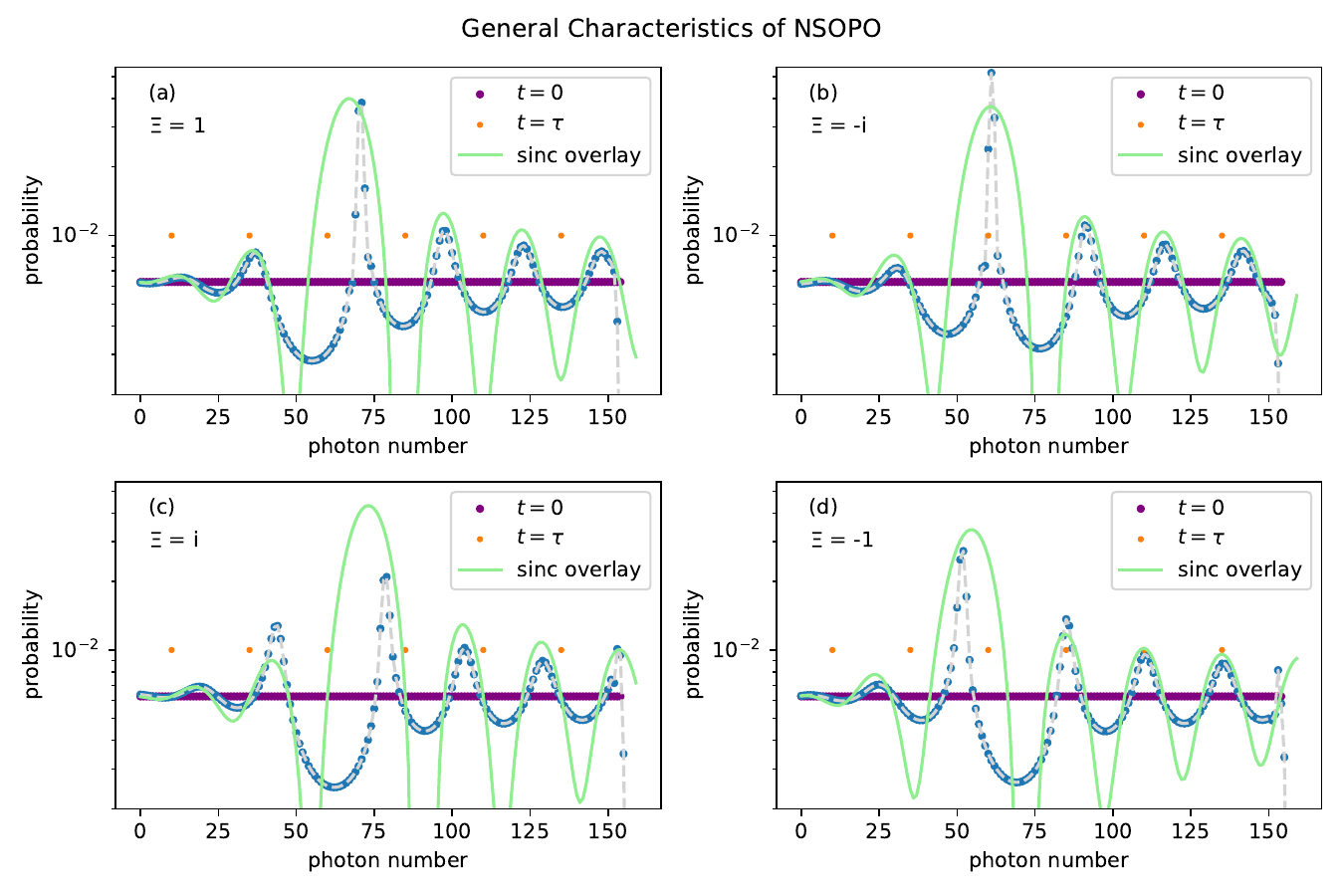}
\end{adjustwidth}
\caption{General characteristics of NSOPO. In each plot the state is initially a uniform real valued distribution. The plots differ only in the phase of the NSOPO complex squeezing parameters $\Xi$. For all plots, the gouge center $n'=60$, the Kerr detuning $\gamma = \pi$, and the evolution time $\tau = 0.08$. Orange dots are spaced by the predicted peak spacing, $s = \frac{2\pi}{\gamma \tau}$, and aligned with the shaping center. The green curve shows a first order approximation of shaping that follows the peak positions and magnitude closely. The blue curve is the shaped photon probability distribution resulting from an in initial uniform distribution. \label{fig1}}
\end{figure}

The time dependent Schr\"odinger equation corresponding to the interaction Hamiltonian, Equation \eqref{HI}, is numerically evolved with a Crank-Nicolson method, which preserves unitarity.  Figure {\ref{fig1}} reveals the characteristic behavior of NSOPO. It shows that probability amplitude tends to coalesce around number states separated by a spacing

\begin{equation}\label{s}
s = \frac{2\pi}{\gamma t}
\end{equation}

where states see constructive interference from two mode squeezing. These peaks rapidly fall off in amplitude when near the shape center, but fall off more slowly when away from the shape center maintaining a wide tail in their height distribution. The peak heights increase monotonically with $|\Xi|$ until reaching a maximum, after which the peak begins to wind itself and develops a complicated structure.

The peak nearest the shape center deviates from periodic spacing, consequently shifting the pattern for states with higher photon number. The location of the central peak depends on the phase of the NSOPO squeezing parameter $\Xi$. For $\Xi = -i$, the peak is centered on the shape center. For $\Xi = i$, there is a deep trough at the shape center with nearest peaks equidistant to the left and right. For $\Xi = 1(-1)$, the nearest peak and trough are to the right and left (left and right) of the shape center. The plots in Figure \ref{fig1} underlay the function

\begin{equation}
\frac{1}{N}\Big[\text{sinc}\Big(\gamma (n'-m)\tau + arg(\Xi)+\pi/2\Big)(m+1)\tau+1\Big]
\end{equation}

which approximately replicates the height and location of each peak, breaking accuracy at the central peak. This function is obtained by integrating the Hamiltonian over time $\tau$ and acts as a more tractable guideline for engineering states with shaping.

%\iffalse

\begin{figure}[H]
\begin{adjustwidth}{-1cm}{-1cm}
\centering
\includegraphics[width=17.5cm]{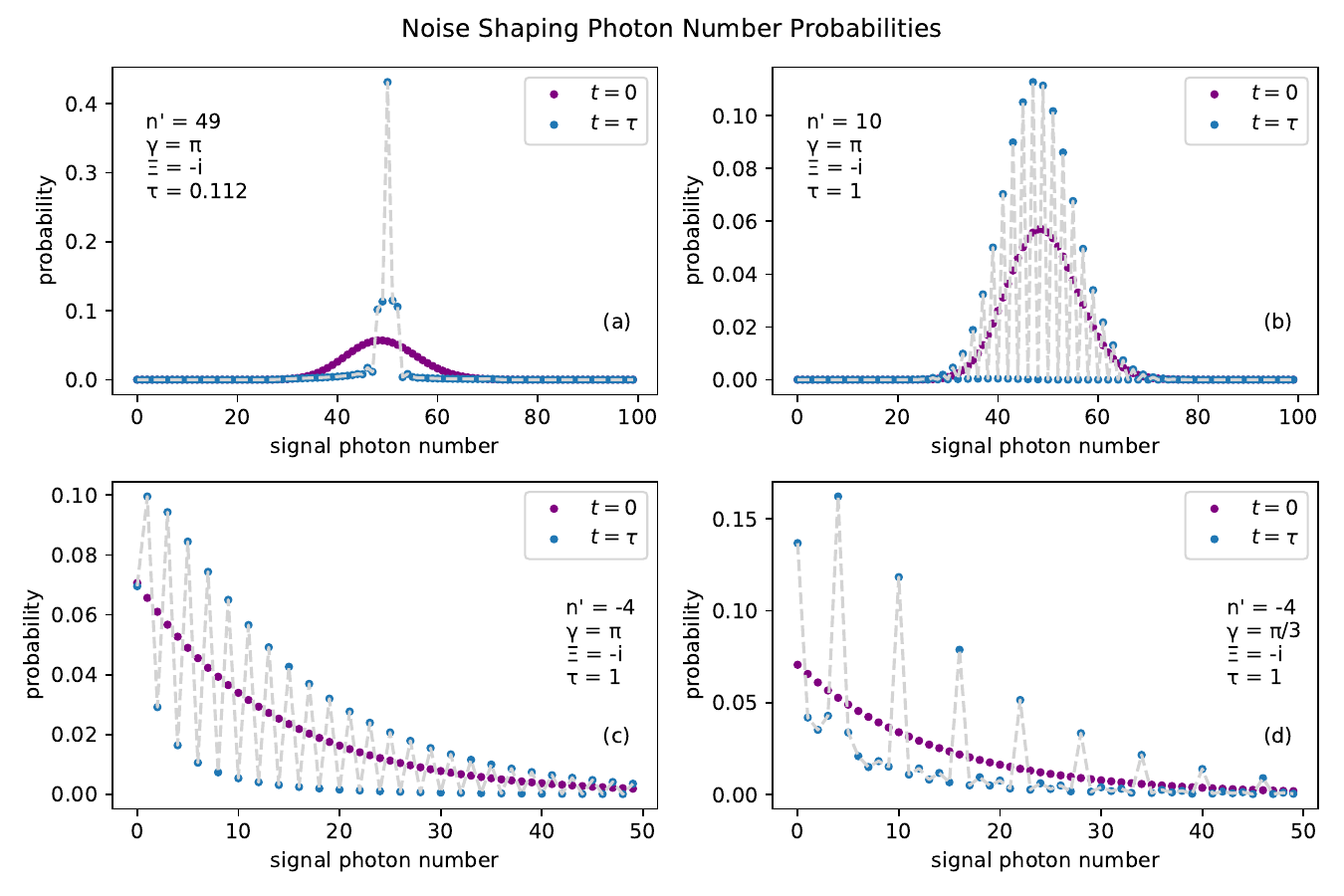}
\end{adjustwidth}
\caption{Examples of photon probability shaping offered by NSOPO.  (a) An approximate Fock state of $n = 52$ accomplished by placing the shaping center on the peak of an initial coherent state with amplitude $\alpha_{s,i} = 7$. (b) A precision shaping of the same coherent state.  Parameter $\gamma t$ is chosen such that peak spacing $s$ is two photons. (c) A two mode squeezed vacuum state with $s = 2$. (d) A two mode squeezed vacuum with $s = 6$.  \label{fig2}}
\end{figure}

These shaping principles can be applied in several ways. Setting $arg(\Xi)= -\pi/2$ such that the shape center is aligned with the peak of a state and the peak spacing $s$ to be greater than the width of the state, the photon number probability will coalesce into an approximate Fock state. Plot (a) in Figure \ref{fig2} illustrates this process, generating a state with signal photon number probabilities $P(n=52) \approx 45\%$, $P(n= 54, 53, 51, 50) \approx 11\%$. Alternatively, if the shape center is set far from the concentration of initial state probabilities, clean periodic oscillations in photon number probabilities, with period $s$, are introduced into the state.  Plots (b-d) in Figure \ref{fig2} illustrate this process. (b) shows precision shaping of a coherent state with $s=2$. (c) is the same as (b) but acting on a two-mode squeezed vacuum. (d) shows shaping with $s=6$.

%\iffalse
%\fi

In tandem with photon number probabilities, NSOPO offers control over states in optical phase space. The following demonstrates an interesting case of this control; generating a family of displaced Schr\"odinger cats with a single operation. Recall that, in the Schr\"odinger picture, the modes experience self phase modulation and cross phase modulation via the Kerr effect. Introducing the condition $G_\Sigma = 0$ into Equation \eqref{H0}, we see that the term of the Hamiltonian

\begin{equation}
\begin{split}
     H^{Kerr} = -\hbar\big(g_{ss}N_s^2 + g_{ii}N_i^2 + g_{is}N_iN_s\big)
\end{split}\;,
\end{equation}

generates non-linear dynamics for the signal and idler modes.
Noting that $N$ is an integer, we see that states of a single mode are periodic in time $t$ when each Kerr frequency satisfies $gt = 2\pi m$, with $m$ an integer. For simple rational values of $m$, the evolution maps coherent states to generalized coherent states \cite{yurke86.01}.

We extend this argument to the two mode squeezed state, recalling $N_i = N_s$. We choose a preferred final state by selecting $m$ to satisfy

\begin{equation}
    (g_{ss} + g_{ii}+ g_{si})t = 2\pi m = (\gamma/2 + g_{\Sigma\Sigma})t\;.\label{closed orbit}
\end{equation}

\begin{figure}[H]
\begin{adjustwidth}{-1cm}{-6cm}
\centering
\includegraphics[width=16cm]{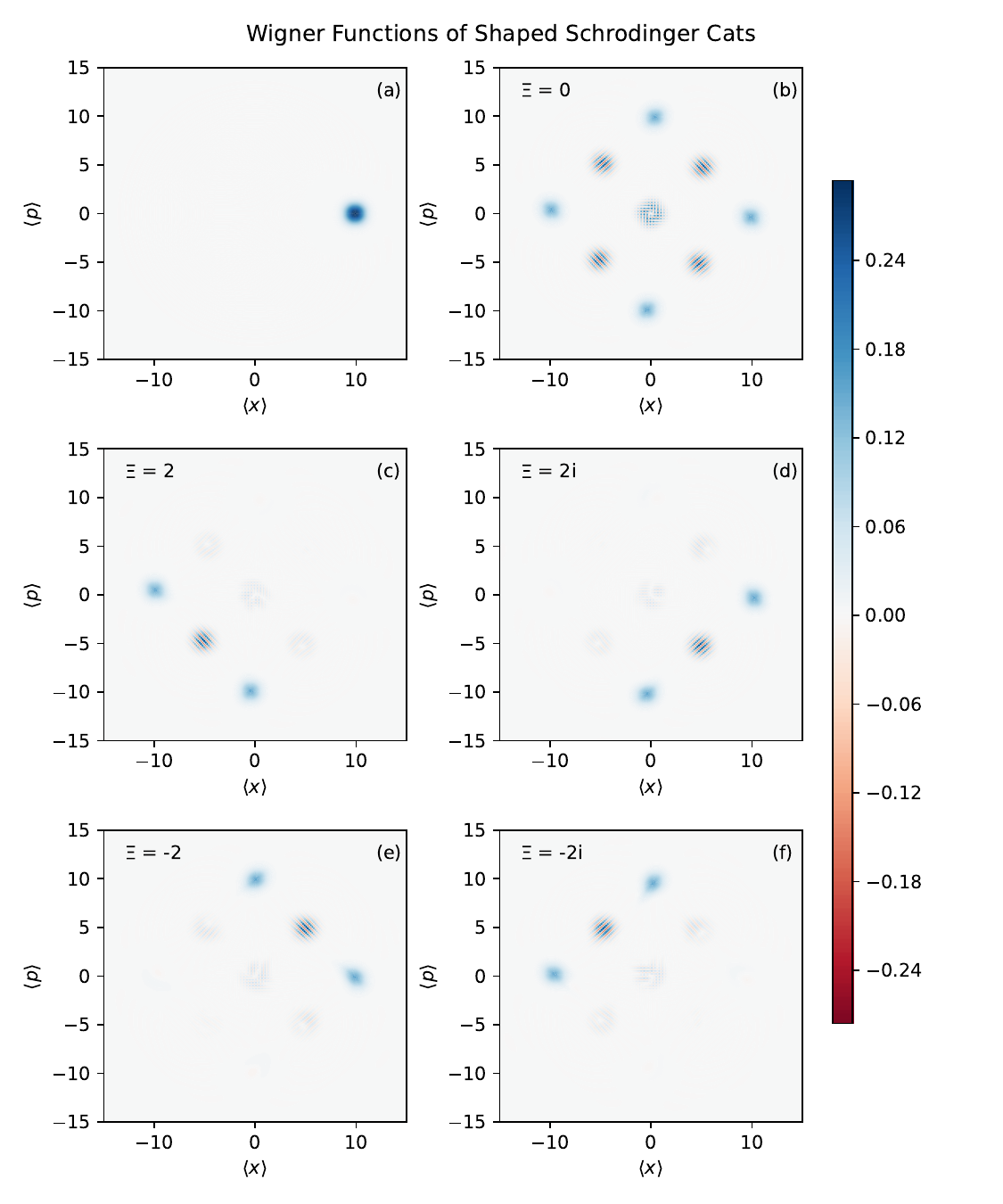}
\end{adjustwidth}
\caption{(a) Wigner function of an initial coherent state of amplitude $\alpha = 7$.  (b) The coherent state subject to Kerr rotation without any NSOPO. (c-f) The state shaped with NSOPO complex squeezing parameter $\pm2, \pm2i$. In all plots, blue regions are positive valued and red regions are negative. Every state is shaped with parameters $\gamma = \pi/2, \tau = 1, n' = 8$. As the phase of $\Xi$ is varied by increments of $\pi/2$, number selective shaping changes predictably.\label{fig3}}
\end{figure}

Equation \eqref{s} provides a similar constraint regarding NSOPO peak spacing. If we suppose $g_{\Sigma\Sigma} = 0$, then $(g_{ss} + g_{ii}+ g_{si}) = \gamma/2$ making Equation \eqref{closed orbit} and Equation \eqref{s} the same constraint with $2m = 1/s$. Setting $s=4$ we obtain a four-phase Schr\"odinger cat state. Figure \ref{fig3} shows the action of NSOPO on a coherent state, (a), that is split into a generalized four-phase Schr\"odinger cat state, (b). Depending on the phase $arg(\Xi)$, NSOPO restricts the amplitude of the cat to two of four phases, shown in (c-f). The axes in each plot are independent phase space quadratures of the displacement field. Not shown in Figure \ref{fig3}, shifting the shape center by one rotates the action of NSOPO among the operations shown in (c-f). Namely, under $n'\rightarrow n'+1$, the plots map to each other according to $ c,d,e,f \rightarrow f,c,d,e$. This behavior alludes efficient quantum logic gates for macroscopic quantum states, though further investigation is needed to understand the range of this capability.

%\iffalse
%\fi

\begin{figure}[H]
\begin{adjustwidth}{-1cm}{-1cm}
\centering
\includegraphics[width=17.5cm]{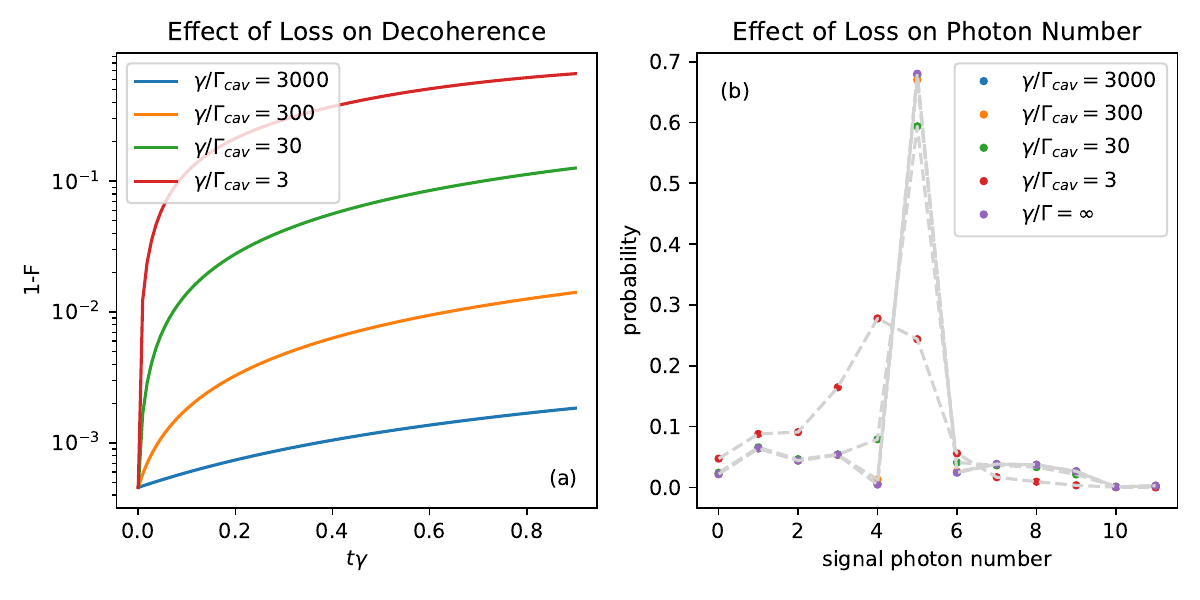}
\end{adjustwidth}
\caption{Both plots present data from a simulation generating an approximate N=5 Fock state in the presence of incoherent loss. The state is evolved from an initial two mode squeezed coherent state of amplitude $\alpha = 2$. The Hamiltonian parameters are $|\Xi|\tau = .3$, $\gamma \tau = 0.9$ and $n' = 4$. Plot (a) shows the growth of quantum infidelity $1-F$ with respect to reciprocal NSOPO peak spacing $\gamma t = \frac{2\pi}{s}$. Each curve represents a different value of the strength to loss ratio $\gamma/\Gamma_{cav}$ where $\gamma$ is the the adjacent state detuning and $\Gamma_{cav}$ is the decay rate of the cavity at the signal and idler frequencies. Plot (b) shows the effect of loss on corresponding photon number probabilities. \label{loss}}
\end{figure}

Figure {\ref{loss}} shows the effect of incoherent loss on the generation of an approximate Fock state, similar to that shown in Figure {\ref{fig2}} (a). In this instance, an initial product state $|\psi\rangle$ of a signal/idler two-mode coherent state with amplitude $\alpha_{s,i} = 2$ and a large coherent pump state is evolved according to the local Lindblad master equation
\begin{equation}
\begin{split}
\frac{d\rho(t)}{dt} &= \frac{1}{i\hbar}[H^{(1)} + H^{(2)} + H^{(3)}, \rho(t)] \\&+ \frac{\Gamma_{cav}}{2}\Big(2a_s\rho(t)a^\dagger_s - \rho(t)a^\dagger_s a - a_s^\dagger a_s \rho(t) + 2a_i\rho(t)a^\dagger_i - \rho(t)a^\dagger_i a - a_i^\dagger a_i \rho(t)\Big)
\end{split}
\end{equation}
where $\rho(0) = |\psi\rangle\langle \psi|$ and $\Gamma_{cav}$ represents the incoherent decay rate of the cavity at the signal and idler frequencies. 

To characterize the viability of the system for generating useful and pure quantum states, we introduce a parameter that is similar in spirit to the cooperativity \mbox{\cite{kroez23.01}}. Cooperativity is defined for light-matter systems with specific quantum models of the material and quantifies the ratio of the light-matter interaction rate to the loss rate of the system. Our theory is derived from a constitutive relation, Equation {\eqref{eq:PExpandDandE}}, so does not offer the detail required to provide the exact cooperativity. However the strength to loss ratio $\gamma/\Gamma_{cav}$ of the adjacent state detuning $\gamma$, i.e. the Kerr frequency of the signal-idler state, and the incoherent decay rate of the cavity at the signal and idler frequencies $\Gamma_{cav}$, serves a similar role.

Plot (a) shows the growth of quantum infidelity $1-F$ with respect to reciprocal NSOPO peak spacing $\gamma t = \frac{2\pi}{s}$. Each curve represents a different value of the strength to loss ratio $\gamma/\Gamma_{cav}$. For $\gamma t <1$, infidelity scales linearly with $\gamma t$. For $|\Xi |t = .3$, $n' = 4$, and $\gamma t = 0.9$, we find the numerical result 
\begin{equation}\label{inf}
\Gamma_{cav}/\gamma \approx \frac{1-F}{3}\;.
\end{equation}
This inverse proportionality between strength to loss ratio and infidelity is substantial. Interestingly, generating approximate Fock states of higher photon number require a smaller $\gamma t$ such that the spacing is larger, thus high Fock state generation demands less of the strength to loss ratio for the same quantum fidelity.

\subsection{Pump Dephasing} \label{Purity}
A key challenge to overcome for preserving the quantum coherence of noise shaped states of light is to mitigate the dephasing effects of the pump beam. The two criteria to satisfy are a strong pump beam, quantified below, and a vanishing detuning weight for the pump frequency, $G_\Sigma = 0$. Specifically, we demand that the detuning weights satisfy

\begin{equation}
 G_\Sigma N_\Sigma \approx |2g_{\Sigma \Sigma} - g_{\Sigma s} - g_{\Sigma i}||\alpha_\Sigma|^2 \ll G_sN_s + G_iN_i = \gamma N_s\;. \label{smallG}
\end{equation}

A vanishing pump detuning weight ensures that the desired OPO detuning $\Delta$ is independent of pump photon number, rendering NSOPO {\it not} number selective for the pump field. This condition is important because a pump photon number selective process implies that far fewer states in the entire Hilbert space contribute to OPO, dramatically decreasing the efficiency of the process. Furthermore, if the interaction were number selective for the pump, the state of the pump field would vary dramatically for different possible interaction histories, resulting in strong dephasing.

When NSOPO is not number selective for the pump, the OPO interaction is simply a displacement operation, which minimally disturbs the structure of the pump field. To illustrate, consider the state $|\alpha_\Sigma, n_s, n_i\rangle$ time evolved under the OPO interaction $H_I$ with $G_\Sigma = 0$. The result is a superposition of products of shifted coherent states in the sum frequency and incremented number states in the signal and idler. The coherent states are at most shifted by a displacement corresponding to the maximum photon number change of the signal and idler. For large amplitude coherent states, shifts in the coherent states on the scale of shifts in the signal and idler yield minimal deviations in the overlap between the coherent pump states.

To obtain a generic measure of state purity, we note that the histories of Fock state amplitudes are carried along with NSOPO peaks. The pump beam coherent states that are entangled with either peak acquire dephasing contributions from pump beam coherent states with mean shifted by the interval of photon numbers the NSOPO peak has traveled over its evolution. This entanglement between the pump, signal, and idler contributes a dephasing in the signal and idler density matrix. The purity of the signal/idler density matrix, $Tr(\rho_{s/i}^2)$, is well approximated by the squared overlap of the shifted and unshifted pump beam states. Thus, for a pump beam in a coherent state $|\alpha_\sigma\rangle$, signal/idler in initial two mode squeezed coherent state $|\alpha_{s,i}\rangle$, and spacing $s$, the purity measure is

\begin{equation}
\gamma_{purity} = \exp\Big\{-\Big|\frac{\alpha_{s,i}}{4s\alpha_\Sigma}\Big|^2\Big\}\;.\label{purity}
\end{equation}

\subsection{Summary of Constraints and Figures of Merit}\label{summary}
Five conditions must be met in order to realize coherent NSOPO. The following subsections summarize each, roughly ordered from simple to difficult to satisfy.

\begin{enumerate}

\item
In the presence of strong Kerr nonlinearities, exciting a cavity pump field to high intensity requires a broad external pump linewidth. However, a broad external pump will excite superfluous modes in the system. Suppose the pump beam is a broad spectrum coherent state $|\alpha_\Sigma\rangle$. To mitigate unintended interactions from modes above the signal and idler frequency, we restrict the magnitude of the sum mode self phase modulation frequency by Equation \eqref{Nmax} to obtain
\begin{equation}
g_{\Sigma\Sigma} < \frac{\omega_\Sigma}{4|\alpha_\Sigma|^4} \;.\label{blockade}
\end{equation}

Thus, it is essential for $g_{\Sigma \Sigma}$ to be small or negative.
\item
The purity measure Equation \eqref{purity} indicates the constraint

\begin{equation}
\frac{\alpha_{s,i}}{4\alpha_\Sigma} \ll s\;.
\end{equation}

The precision of noise shaping is limited by the relative uncertainty of the signal/idler photon number noise w.r.t. the pump photon number noise and is generally the easy to control.

\item
Numerical simulations place useful values of $\frac{\Xi}{\gamma}$ in the range $0.1$ to $10$. For an order of magnitude estimate of this constraint, consider a case where $g_{ss} = g_{ii} = g_{is}$ and $g_{\Sigma\Sigma} = 0$. Using the definitions of $\gamma$, $\Xi$,and $g$ from equations \eqref{gamma},  \eqref{Xi}, \eqref{Kerrfreq}, \eqref{zeta},

\begin{equation}
\frac{\Xi}{\gamma} = \frac{\Xi}{6g} \approx \frac{\zeta_\Sigma \Gamma^{(2)}\alpha_\Sigma^*}{12\zeta_s\zeta_i\Gamma^{(3)}}\;.
\end{equation}

The quantity $\zeta$ is the single photon displacement field amplitude in Equation \eqref{zeta}. Careful use of Equation {\eqref{gamma2chi}} relates the displacement field susceptibilities to the more commonly measured electric field susceptibilities. Expanding the ratio with the definitions of each frequency yields

\begin{equation}\label{c4}
\begin{split}
\frac{\Xi}{\gamma} &\approx \sqrt{\frac{LA}{18\hbar\omega}}\sqrt{\frac{cv_{g,\Sigma}n_\Sigma^3}{v_{g,s}v_{g,i}n_s^3n_i^3}}\frac{\Gamma^{(2)}\alpha_\Sigma^*}{\Gamma^{(3)}} \\& \approx \sqrt{\frac{LA}{18\hbar\omega}}\sqrt{\frac{cv_{g,\Sigma}}{v_{g,s}v_{g,i}}} \sqrt{\frac{1}{n_\Sigma n_s n_i}}\frac{\chi^{(2)}\alpha_\Sigma^*}{\chi^{(3)}} \in [0.1, 10]
\end{split}
\end{equation}

This criterion is straightforward to satisfy in real materials. For however large the magnitude of $\chi^{(3)}$ is, the strength of OPO can be increased with the intensity of the pump beam and the quantization volume.

\item
As discussed in Subsection \ref{Purity} above, shaping maintains quantum coherence if the pump field detuning weight vanishes, $G_\Sigma = 0$. Equation \eqref{smallG} provides a qualitative expression of the requisite smallness of $G_\Sigma$. The upshot is that greater pump power requires higher precision in the proximity of $G_\Sigma$ to zero. This condition is easiest to satisfy if Kerr frequencies related to the pump are independently small, which is challenging if the other Kerr frequencies are to be incredibly large. Appendix {\ref{app}} discusses a class of systems that can satisfy exactly this scenario by generating nonlinearities on resonance with the signal and idler, and off resonance with the pump.

\item
To apply the theory developed in this paper, all shaping should occur on a time scale $\tau$ shorter than that defined by the acceptable infidelity $1-F_{acc}$ and desired spacing according to
\begin{equation}
\frac{\Gamma}{\gamma}\frac{2\pi}{s} = \frac{\tau}{\tau_{cav}} < \frac{1-F_{acc}}{s} \;.
\end{equation}
where $\tau_{cav} = 1/\Gamma_{cav}$ is the cavity decay time. This constrains the lower bound on the Kerr frequencies $g_{ss}, g_{ii}, g_{si}$. Let us assume $g_{ss} = g_{ii} = g_{si} = g$ and $g_{\Sigma,\Sigma} = 0$ for simplicity. Restating the numerical result for the high fidelity regime Equation {\eqref{inf}}, but substituting in the cavity lifetime $\tau_{cav} = \frac{2Q}{\omega}$ we obtain
\begin{equation*}
\frac{3}{1-F} \approx \frac{\gamma}{\Gamma_{cav}} = \frac{12gQ}{\omega} \;.
\end{equation*}

Substituting $\zeta$ from Equation {\eqref{zeta}} into the definition of the Kerr frequencies $g$ from Equation {\eqref{Kerrfreq}}, and relating $\Gamma^{(3)}$ to $\chi^{(3)}$ with Equation {\eqref{gamma2chi}}, we find the strength to loss ratio is
\begin{equation}\label{c5}
\frac{\gamma}{\Gamma_{cav}} = 72\pi Q\frac{\hbar \omega}{V}\frac{v_g^2}{c^2}\frac{\chi^{(3)}}{n^2}
\end{equation}

Finally, substituting in the classical form of the nonlinear refractive index $n_2 = \frac {24\pi^2\chi^{(3)}} {c n^2}$ we obtain a figure of merit $\mathcal {F}$ for NSOPO that is proportional to the strength to loss ratio, 
\begin{equation}\label{FOM}
\frac{1}{1-F} \approx \frac{Q\hbar \omega v_g^2 n_2}{\pi Vc} \equiv \mathcal {F}
\end{equation}

This figure of merit commands our attention; it is the most challenging constraint to practically satisfy among this list. Candidate systems will have large $\mathcal{F}$ at the signal and idler frequencies and small $\mathcal{F}$ at the sum frequency. In Appendix {\ref{app}} we show that compliance is likely attainable by addressing the state of the art for achieving strong nonlinearities with minimal loss and controllable group velocity and showing that these systems perform well within the boundary of the constraints.

\end{enumerate}

\subsection{Control Parameters}
The shape center $n'$ is controlled by modulating the linear refractive index of the material, therefore the static detuning parameter $\delta$, in time. To understand the extent of modulation we may ask, what are the possible rates of change for the shape center with respect to refractive index shifts? Recall the expression for $\delta$, \eqref{KerrDetuneings}, contains the frequency difference $\omega_{\Sigma} - \omega{s} - \omega_i$. In the simplest case, the frequencies are related to the fixed wavenumbers of each mode by

\begin{equation}
 \omega = \frac{kc}{n} \;.
\end{equation}

Where $\frac{d\omega}{dn}$ is maximal, so is $\frac{d\delta}{dn}$.  Therefore it is desirable for the refractive index of each mode to be near zero. This conclusion is agreeable with the figures of merit derived above. One can change the refractive index of the material by, for example, introducing a static field, or by utilizing acoustic oscillations in the cavity material.

The NSOPO squeezing parameter, \eqref{Xi}, is primarily controlled by modulating the amplitude and phase of the pump beam. In practice one should be aware that changing the shape center affects the strength of NSOPO in three ways.
\begin{enumerate}
\item Equation \eqref{HI} shows that OPO strength is scaled by $(m+1)$, where $m$ is the signal photon number of the state being acting upon. Thus the OPO strength is scaled by the shape center.
\item  In some materials, e.g. organic materials, changing a static field to modulate the refractive index will reorient molecules affecting $\Gamma^{(2)}$.
\item Static fields will also introduce a $\Gamma^{(3)}$ contribution to OPO strength $\Xi$.
\end{enumerate}
Ultimately, determining the parameters for preparing a particular state will be a matter of empirical tuning guided by rough numerical estimates.

A final note on state control: if the static field can be modulated on a time scale faster than the state preparation time $\tau$, more complicated shaping of the field noise statistics can be accomplished, beyond fixed shaping parameters.

\section{Conclusion}

We have proposed a novel method for the deterministic manipulation of quantum states of light.  The underlying theoretical tools required for its implementation are developed and the physical constraints for its practical implementation described. The method utilizes strong Kerr nonlinearities to introduce photon number dependence into the frequency matching condition for optical parametric oscillation, rendering the OPO interaction photon number selective. Number selective interactions expand the capabilities of current CV quantum optical state control and preparation techniques by introducing targeted deterministic manipulations to subspaces of a light field's Fock space.

Simulations indicate the generation of highly nonclassical quantum states of light. We establish five necessary conditions to realize NSOPO in practice, which will aid the search for candidate materials and optical configurations. The constraints are consistent with a common figure of merit, $\mathcal{F}$, expressed in Equation \eqref{FOM}. The best materials for NSOPO will have large $\mathcal{F}$ at the signal and idler frequencies, and small $\mathcal{F}$ at the sum frequency. The constraints can likely be realized in specialized systems that are driven into electronically induced transparency and spontaneously generated coherence. 

Future research will aim to realize the requisite constraints in detailed models of highly resonant cavity light coupled with matter systems, with special attention paid to dark resonances. From specific models, the approach derived in this paper will be adapted to accommodate Hamiltonians beyond the simple cubic and quartic interactions treated above. These new models will be used to guide experimental design and to search for novel sculpted states. Future research will also investigate generation of entanglement between two uncorrelated signal and idler states, and study how noise shaping can be generalized to pulses of light, in particular to solitons in optical resonators with attention paid to mitigating the dephasing effects of quantum soliton propagation.

\vspace{6pt}

\noindent\textbf{Acknowledgments:} We thank Ned and Nancy Wogman for the Wogman fellowship, the WSU CAS Graduate Research Scholarship, and the donors to the Innovative Physics Fund and the Department of Physics for support.\\

\section*{Appendix}\label{app}
The present theoretical paper describes how one might achieve deterministic shaping of a Fock state.  This appendix gives a plausibility argument that the required figure of merit is achievable.  However, meeting the constraints discussed in section {\ref{summary}} may require new materials and device architectures.  Here we individually consider the measured properties of known material systems to estimate a reasonable range of the figure of merit.

EIT is an example of a process that exhibits the requisite properties for NSOPO. Schmidt and Imamoglu \mbox{\cite{imamo97.01}} showed that a four level system can be configured to obtain a giant cross phase modulation strength. With the additional presence of spontaneously generated coherence (SGC), which can be accomplished in a five level system \mbox{\cite{bang24.01}}, the EIT window can be shifted such that the Kerr nonlinearity is maximal when absorption is near zero. Furthermore, the five level system proposed offers direct control over the group velocity, which ranges from highly sub- to super-luminal and with photon number dependent behavior \mbox{\cite{doai20.01}}.

Obtaining giant nonlinearities in near resonant systems with EIT gives us two other desirable features for NSOPO. The peaks of the linear and nonlinear susceptibilities are sufficiently sharp near resonance to make FWM minimal due to a rapidly increasing phase mismatch and rapidly decaying interaction strength. In addition, off resonant nonlinearities will be comparatively weak. Such a system can be tuned so that the sum frequency is off resonance with the material while the signal and idler are on resonance. Then Constraint {\eqref{smallG}} is satisfied by virtue of small Kerr nonlinearities at the pump frequency and FWM can be reliably neglected for all modes.

Bang and Doai, \mbox{\cite{bang24.01, doai20.01}}, report values of self and/or cross phase modulating nonlinear index in a gas of {${}^{85}$}Rb with density {$10^8$} atoms/cm{$^3$} on the order of
\begin{equation}
n_2 = \frac {24\pi^2\chi^{(3)}} {c n^2} \approx 5\times 10^{-13}\frac {s\cdot cm^2}{erg}
\end{equation}

In principle, the atom density can be made orders of magnitude greater in a solid state device. With modest concentrations of the active molecule on the order of $10^{18}$ molecules/cm$^3$ \mbox{\cite{singer88.01}}, the third order susceptibility for non-interacting molecules is enhanced in proportion to the molecular density, increasing the nonlinear index to 
\begin{equation}
n_2 \approx 5\times 10^{-3}\frac {s\cdot cm^2}{erg}\;.
\end{equation}
Suppose the signal and idler wavelengths are on the order of $1\mu m$, the resonator cross-sectional area is $1\mu m\times 1\mu m$, the group velocity of the signal and idler are both $c$.  Then the figure of merit $\mathcal{F}$ is to within an order of magnitude given by
\begin{equation}
\mathcal{F} = \frac{Q}{L}10^{3}\;.
\end{equation}
For a length on the order of $10\lambda$ and cavity quality factor on the order $10^4$ \mbox{\cite{cheng17.01, ling11.01}}, the infidelity $\mathcal{F}^{-1}$ is $10^{-10}$. This scenario is the upper boundary of feasible materials of this type, so we do not expect to reach it in practice, but it offers several orders of magnitude of leeway for trade-offs in designing materials and devices that shape photon statistics.

While this rough calculation shows the plausibility of deterministic shaping given known materials properties, an important question remains as to the plausibility of making a solid state waveguide device. Realizing the benefits of EIT and SGC in a solid state waveguide device would require a material that in analogy is made with atoms that do not strongly interact with each other. A possible material system is a dye-doped polymer, which can be fabricated into waveguides, and whose linear and nonlinear-optical properties can be controlled with atom/molecule doping.

One way to incorporate the active atom/molecule into a solid is to use a polymer, sol-gel host, or other host -- which would lead to higher dopant concentration, thus a larger nonlinearity \mbox{\cite{dirk90.01}}. Such materials can be made into thin films and fibers \mbox{\cite{welke98.01}} and photolithographic techniques can be used to make waveguide resonators \mbox{\cite{zheng12.01}}. Interactions between the host material and the atom/molecule can be minimized by using molecular pincers --  called chelates -- to isolate them. Chelated rare-earth atoms are commonly used in amplifiers \mbox{\cite{koepp97.01}}. The ability to get higher densities of active molecules to increase nonlinearity, the flexibility of choosing molecules for their spectral properties, the  shielding from the environment to mitigate de-phasing through chelation, and the reduction of inhomogeneous broadening by cooling the material all provide the means for optimization.

Each of these approaches presents challenges, but designing a device is far beyond the scope of this paper. However, the numbers suggest that making a device for deterministic shaping of quantum light statistics is possible and our work defines the material requirements that will guide future experimental efforts.

\bibliographystyle{ieeetr}
\bibliography{NSOPOBib}

\end{document}